\documentclass[showpacs,twocolumn,prl,amssymb,amsmath,nobibnotes,aps]{revtex4}
\usepackage{bm}

\newcommand{\as}{\tilde{a}}
\newcommand{\oh}{{\frac{1}{2}}}

\newcommand{\qv}{{\bf q}}
\newcommand{\kv}{{\bf k}}

\newcommand{\be}{\begin{equation}}
\newcommand{\ee}{\end{equation}}
\newcommand{\bea}{\begin{eqnarray}}
\newcommand{\eea}{\end{eqnarray}}

\usepackage[pdftex]{graphicx}
\begin{document}

\title{Quench dynamics of a strongly interacting resonant Bose gas}
\author{Xiao Yin}
\author{Leo Radzihovsky}
\affiliation{Department of Physics, University of Colorado,
   Boulder, CO 80309}
\date{\today}

\begin{abstract}
  We explore the dynamics of a Bose gas following its quench to a
  strongly interacting regime near a Feshbach resonance.  Within a
  self-consistent Bogoliubov analysis we find that after the initial
  condensate-quasiparticle Rabi oscillations, at long time scales the
  gas is characterized by a nonequilibrium steady-state momentum
  distribution function, with depletion, condensate density and
  contact that deviate strongly from their corresponding equilibrium
  values. These are in a qualitative agreement with recent experiments
  on $^{85}$Rb by Makotyn, et al. Our analysis also
  suggests that for sufficiently deep quenches close to the resonance
  the nonequilibrium state undergoes a phase transition to a fully
  depleted state, characterized by a vanishing condensate density.
\end{abstract}
\maketitle 

Experimental realizations of trapped degenerate atomic gases coupled
with field-tuned Feshbach resonances (FR) \cite{FRrmp} have led to
studies of quantum states of matter in previously unexplored,
extremely coherent, strongly interacting regimes.  Some of the notable
early successes include a realization of paired s-wave superfluidity
and the corresponding BCS-to-Bose-Einstein condensate (BEC)
crossover\cite{GrimmBCS-BEC,JinBCS-BEC,KetterleBCS-BEC}, phase
transitions driven by species
imbalance\cite{PartridgePolarized,RSpolarized}, and the
superfluid-to-Mott insulator transition in optical
lattices\cite{GreinerSIT}. More recently, much of the attention has
turned to nonequilibrium analogs of these quantum states, made
possible by unmatched high tunability (adiabatic or sudden quench) of
system parameters, such as FR interactions and single-particle (e.g.,
trap and lattice) potentials in atomic gases. Quenched dynamics of FR
fermionic gases have been extensively explored theoretically,
predicting coherent post-quench oscillations \cite{Barankov,AGR} and
topological nonequilibrium steady states and phase
transitions \cite{Foster}. In bosonic gases such quench studies date
back to seminal work on $^{85}$Rb \cite{Donley}, illustrating coherent
Rabi-like oscillations between atomic and molecular
condensates \cite{HollandKokkelmans}.  More recently, oscillations have
also been observed in quasi-two-dimensional bosonic $^{133}$Cs, following shallow
quenches between weakly-repulsive interactions \cite{Chin}, and have stimulated theoretical studies of weak two-dimensional quenches \cite{Natu,Levin}. Given that
resonant bosonic gases are predicted to exhibit atomic-to-molecular
superfluid phase transition (rather than just a fermionic smooth
BCS-BEC crossover) and other interesting
phenomenology \cite{RPWmolecularBEC}, we expect their quenched dynamics
to be even richer, providing further motivation for our study.

Fundamentally, such Bose gases become unstable upon approach to a FR
(where two-particle scattering length $a_s$ diverges) due to a growth
of the three-body loss rate $\gamma_{3}\propto n^2 a_s^4$ relative to
the two-body scattering rate $\gamma_2\propto n^{4/3} a_s^2$. On
general grounds, in the limit of $n a^3_s\gg 1$, these rates are
expected and found \cite{Makotyn} to saturate at an order of Fermi-like energy (energy set by
atom density, which for simplicity we will just call Fermi energy) $\epsilon_n=\hbar^2k_n^2/2m$ ($k_n\equiv n^{1/3}$), exhibiting
universality akin to unitary Fermi
gases \cite{universalHo,Nicolic,Veillette}. However, as was recently
discovered in $^{85}$Rb \cite{Makotyn}, quenches on the molecular
($a_s>0$) side of the resonance, even near the unitarity, the
three-body rate appears to be more than an order of magnitude slower
than the two-body rate (both proportional to Fermi energy, as
expected), thereby opening up a window of time scales for metastable
strongly-interacting nonequilibrium dynamics.

Stimulated by these experiments \cite{Donley,Chin,Makotyn} and taking
the aforementioned slowness of $\gamma_3\ll\gamma_2$ as an empirical
fact, in this report we study the upper-branch effectively repulsive
dynamics of a three-dimensional gas of strongly interacting bosonic
atoms following a {\em deep} detuning quench close to the unitary
point on the molecular side ($a_s > 0$) of the FR. Before turning to
the analysis, we present highlights of our results that show
qualitative agreement with JILA experiments \cite{Makotyn} and discuss the
limits of their validity.

\begin{figure}[htb]
\centering
\includegraphics[width=0.45\textwidth]{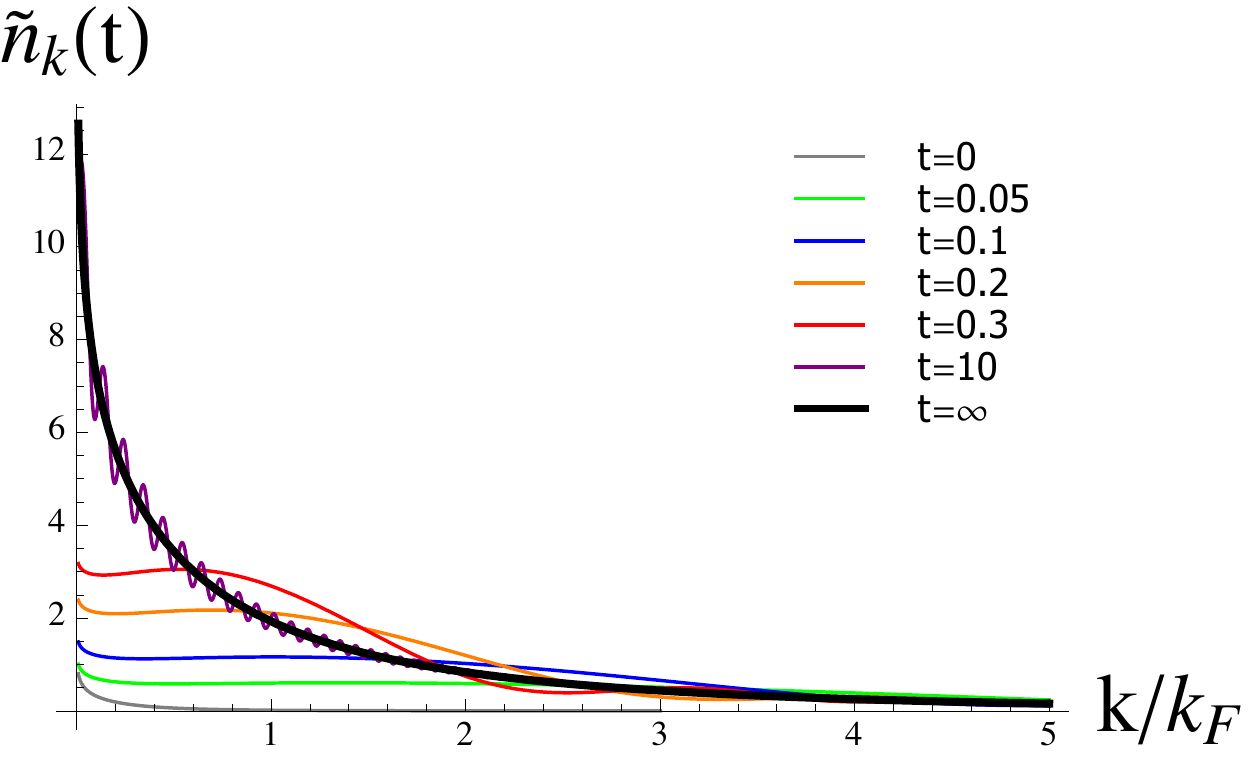}
\caption{(Color online) Time evolution of the (column-density) momentum distribution
  function, $\tilde{n}_{\kv_\perp}(t)\equiv n^{-1/3}V^{-1}\int dk_z
  n_\kv(t)$ following a scattering length quench
  $k_na_0=0.01\rightarrow k_na_f=0.6$ in a resonant Bose gas.}
\label{nk_tFig}
\end{figure}

\begin{figure}[htb]
\centering
\includegraphics[width=0.45\textwidth]{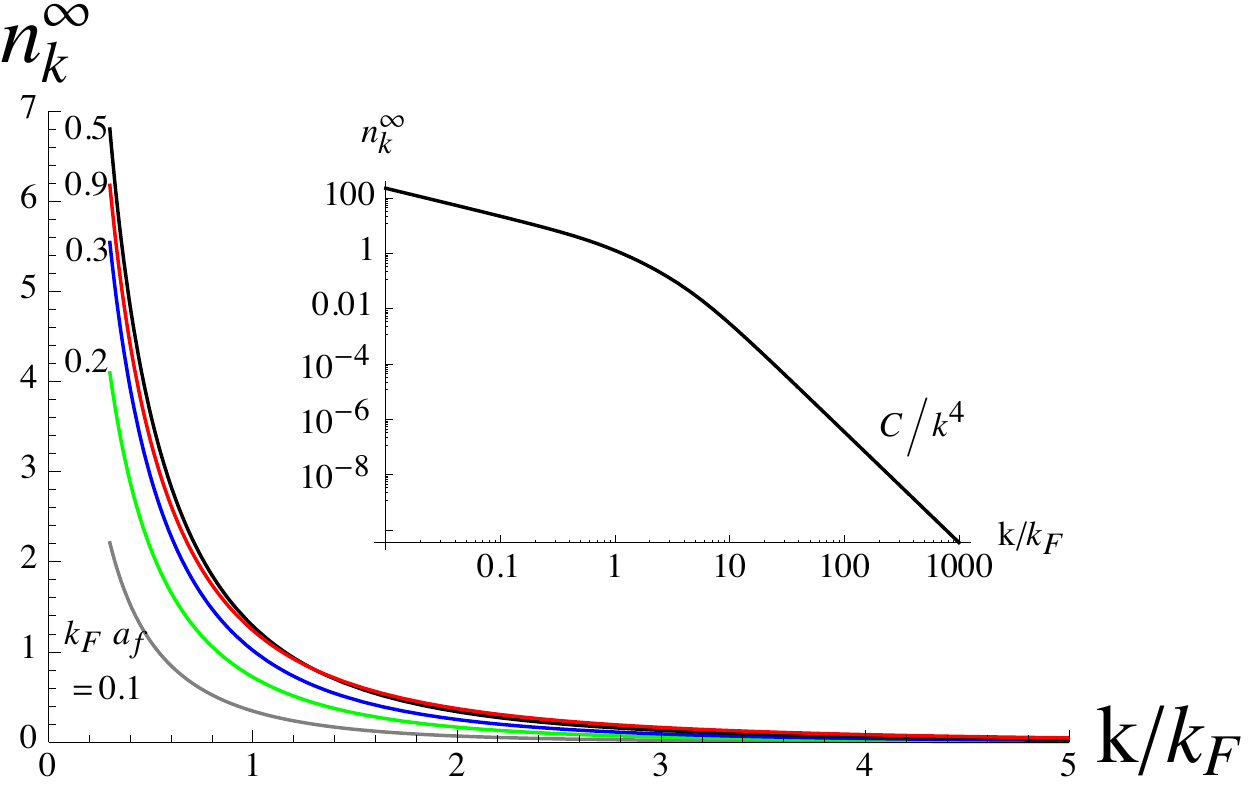}
\caption{(Color online) A nonequilibrium steady-state momentum distribution function
  $n_k^\infty$, approached at long times, following a scattering
  length quench $a_0\rightarrow a_f$ (illustrated for $k_n a_f =
  0.1,0.2,..., 0.9$ and $k_na_0=0.01$) in a resonant Bose gas. The
  inset illustrates the presence of the $1/k^4$ large momentum tail.}
\label{nk_ssFig}
\end{figure}

\begin{figure}[htb]
\centering
\includegraphics[width=0.45\textwidth]{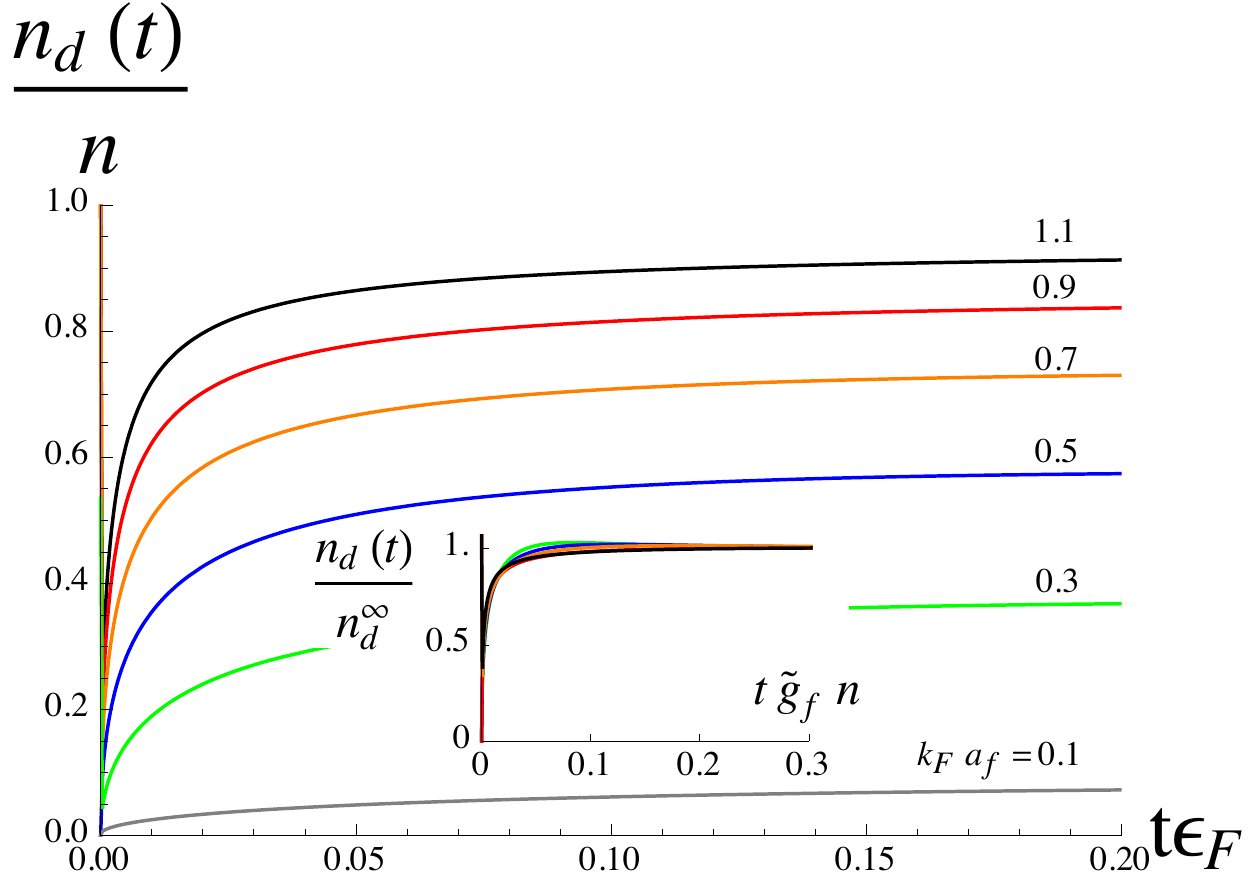}
\caption{(Color online) Time evolution of the condensate depletion, $n_d(t)$ as a
  function of pulse duration $t$, following a scattering length quench
  $a_0\rightarrow a_f$ in a resonant Bose gas. The inset shows an
  approximate collapse of the scaled depletion, with only a weak
  dependence on other parameters.}
\label{ndtFig}
\end{figure}

Following a sudden shift in interaction from $g_0=4\pi a_0/m$ to
$g_f=4\pi a_f/m$ (we take $\hbar=1$ throughout) leaves the system in
an excited state of the shifted Hamiltonian, $H_f$ that leads to a
nontrivial dynamics associated with Rabi-like oscillations between an
atomic condensate and Bogoliubov quasi-particles. Although such
oscillations have indeed been seen in shallow
quenches\cite{Donley,Chin}, they do not appear to have been observed
in deep quenches to unitarity \cite{Makotyn}.  We find that
oscillations at frequencies corresponding to different momenta $k$
decohere on longer time scales, set by the inverse of the Bogoliubov
spectrum. We thus observe that the momentum distribution function,
$n_k(t)=\langle 0^-|a^{\dagger}_k(t) a_k(t)|0^-\rangle$ approaches a
steady state form at a time scale growing as $\sim 1/k^2$ ($\sim 1/k$)
for momenta larger (smaller) than the coherence momentum,
$k_\xi=\sqrt{4\pi a_f n}$. The full distribution function reaches a
nonequilibrium steady state at the longest time scale $R\sqrt{m/(g_f
  n)}$ set by the cloud size, $R$, beyond which it is characterized by
a time-independent momentum distribution function illustrated in
Fig.\ref{nk_ssFig}. Within our self-consistent Bogoliubov theory
$n_k(t\rightarrow\infty)\equiv n_k^\infty$ never fully thermalizes,
although is expected to on physical grounds if Bogoliubov
quasi-particle collisions are taken into account. However, if this
latter rate is significantly slower than the interaction energy (set
by the chemical potential), we would expect our prediction for
$n_k(t)$ (Fig.\ref{nk_tFig})
and its long-time steady-state form (illustrated in Fig.\ref{nk_ssFig})
\begin{eqnarray}
\label{nk_infty}
\hspace{-0.3cm}n_k^\infty
&=&\frac{k^4+8\pi k^2(a_0n+\as_f n + \as_f n_c^\infty) 
+ 64\pi^2 \as_f n_c^\infty n(\as_f+a_0)}
{2\sqrt{k^2(k^2+16\pi\as_f n_c^\infty)(k^2+16\pi a_0 n)
    (k^2+16\pi\as_f n)}}\nonumber\\
&&-\oh
\end{eqnarray}
to capture the nonequilibrium momentum distribution observed in JILA
experiments\cite{Makotyn}.
In Eq.\ref{nk_infty} $\as_f = a_f/\sqrt{1+k_n^2a_f^2}$ is the
effective finite-density scattering length and $n_c^\infty\equiv
n_c(t\rightarrow\infty)$ is the asymptotic steady-state value of the
condensate density, $n_c(t) = n - V^{-1}\sum_{\kv\neq0}n_k(t)$, which
is self-consistently determined by the depletion
$n_d(t)=V^{-1}\sum_{\kv\neq0}n_k(t)$, illustrated in Fig.~\ref{ndtFig}.
After time set by $m/(4\pi\tilde a_f n)$ the depletion saturates at a
nonequilibrium value, $n_d^\infty(k_n a_f)$ (calculated below), which
deviates significantly from the adiabatic depletion, i.e., the
equilibrium value corresponding to the scattering length $a_f$.

Finally, we find that for $k_n a_f > k_n a_{fc}=1.35$, the asymptotic
condensate density is driven to zero; this contrasts with the
equilibrium state, where in three dimensions at $T=0$ the gas is a BEC at arbitrary
strong interactions, $k_n a_s$. Our analysis thus
suggests\cite{commentTransition} that for a sufficiently deep quench,
the system undergoes a nonequilibrium phase transition to a non-BEC
steady state. We conjecture that the nonequilibrium transition is set
by the critical depth of the quench for which the associated
excitation energy, $E_{exc}=\langle 0^-|H_f|0^-\rangle-\langle
0_f|H_f|0_f\rangle$ (where $|0_f\rangle$ is the ground state of $H_f$)
significantly exceeds $\epsilon_n$.

We obtain these results using a self-consistent Bogoliubov theory,
implemented in two steps. First, to ensure that the condensate fraction
$n_c(t)$ remains positive [equivalently, the depletion $n_d(t)$ does
not exceed the total atom number $n$, as it can at strong coupling within
straight Bogoliubov theory], we solve the Heisenberg equations of
motion for the finite momentum quasi-particles using a time-dependent
Bogoliubov Hamiltonian, where $n_c(t)$ appears as a self-consistently
determined function. This approximation is the bosonic analog of the
self-consistently determined BCS gap function in fermionic
systems\cite{Barankov,AGR} and for a uniform state is equivalent to
solving the Gross-Pitaevskii equation for the condensate $\Phi_0$ in
conjunction with Heisenberg equations for the finite momentum
excitations $a_\kv$. Our second ``beyond-Bogoliubov'' approximation is
the replacement of the scattering length $a_f$ by the density
dependent scattering amplitude $|f(k_n,a_f)|=
a_f/\sqrt{1+k_n^2a_f^2}\equiv \as_f$,
This qualitatively captures the crossover from the two-atom regime,
$a_f\ll n^{-1/3}$ to a finite density limit, when $a_f$ reaches
inter-particle spacing and the scattering amplitude saturates at $\sim
k_n^{-1}$. While the details of the crossover function are \textit{ad hock},
our qualitative predictions are insensitive to these details and
depend on the limiting values of the two regimes.

After the $a_0\rightarrow a_f$ quench the atomic resonant gas with a
free dispersion $\epsilon_k = k^2/2m$ is governed by a single-channel
bosonic Hamiltonian $H_f=\sum_\kv\epsilon_k a_\kv^\dagger a_\kv
+\frac{g_f}{2V}\sum_{\kv_1,\kv_2,\qv}a_{\kv_1}^\dagger
a_{-\kv_1+\qv}^\dagger a_{\kv_2} a_{-\kv_2+\qv}$, with the final
interaction parameter, $g_f$ and corresponding scattering length,
$a_f$, tunable via a magnetic field near a Feshbach
resonance. Motivated by the experiments\cite{Makotyn}, we focus on the
initial states and their subsequent evolution, which, although possibly
strongly depleted and time-dependent, are confined to a
well-established condensate (focusing for simplicity on periodic
boundary conditions without a trap). This allows us to make progress
in treating the resonant interactions, by expanding in finite-momentum
quasi-particle fluctuations about a macroscopically occupied $\kv=0$
state, and thereby to reduce the Hamiltonian to the quadratic form,
$H_f(t)\approx \frac{g_f}{2V}N^2 - \sum_\kv(\epsilon_\kv + g_f n_c(t))
+ H^B_f$(t), where
\begin{eqnarray}
\label{HB}
\hspace{-0.15cm}
H^B_f&=&\oh\sum_{\kv\neq 0} 
\begin{pmatrix}
  a_\kv^\dagger & a_{-\kv}\\
\end{pmatrix}
\hspace{-0.15cm}
\begin{pmatrix}
  \epsilon_k + g_f n_c(t)& g_f n_c(t)\\
  g_f n_c(t) & \epsilon_k+ g_f n_c(t)\\
\end{pmatrix}
\hspace{-0.15cm}
\begin{pmatrix}
  a_\kv\\
  a_{-\kv}^\dagger\\
\end{pmatrix}\nonumber\\
&\equiv&\oh\sum_{\kv\neq 0} 
{\vec\Phi}^\dagger_\kv(t)\cdot \hat{h}_k(t)
\cdot{\vec\Phi}_\kv(t),
\end{eqnarray}
with a new ingredient that the time-dependent condensate density is
self-consistently determined by $n_c(t) = n - \frac{1}{V}\sum_{\kv\neq
  0} \langle 0^-|a_\kv^\dagger(t) a_\kv(t)|0^-\rangle$ in the initial,
pre-quench state $|0^-\rangle$ at $t=0^-$.  Focussing on zero
temperature, we take state $|0^-\rangle$ to be the vacuum with respect
to the quasi-particles $\alpha_\kv$, that diagonalize the initial
Hamiltonian, $H_0^B=\sum_\kv E_\kv^0\alpha_\kv^\dagger\alpha_\kv +
const.$, characterized by a pre-quench scattering length, $a_0$.

The corresponding Heisenberg equation of motion
$i\sigma_z\partial_t\vec{\Phi}_\kv(t)=\hat{h}_k(t)
\cdot\vec{\Phi}_\kv(t)$ for ${\vec\Phi}_\kv(t)=(a_\kv(t),
a_{-\kv}^\dagger(t))$ is conveniently encoded in terms of a
time-dependent Bogoliubov transformation $U_\kv(t)$,
$\Phi_\kv(t)=U_\kv(t)\Psi_\kv$, where
$\Psi_\kv=(\beta_\kv,\beta^{\dagger}_{-\kv})$ are time-independent
bosonic reference operators (ensured by
$|u_\kv(t)|^2-|v_\kv(t)|^2=1$), that diagonalize the Hamiltonian at
the initial time $t=0^+$ after the quench, $H_f^B(0^+)=\sum_\kv
E_\kv^f(0^+)\beta_\kv^\dagger\beta_\kv + const.$. Equivalently,
$U^\dagger_\kv(0^+) h_f(0^+)
U_\kv(0^+)=E_f(0^+)=\sqrt{\epsilon_k^2+2g_fn_c(0^+)\epsilon_k}$,
fixing the initial condition
$u_\kv(0^+)=\sqrt{\oh(\frac{\epsilon_k+g_fn_c(0^+)}{E_f(0^+)} + 1)}$, and
$v_\kv(0^+)=-\sqrt{\oh(\frac{\epsilon_k+g_fn_c(0^+)}{E_f(0^+)} - 1)}$
for spinor $\psi_\kv(t)\equiv (u_\kv(t),v_\kv(t))$, which evolves
according to
\begin{eqnarray}
\label{Heom}
i\sigma_z\partial_t\vec{\psi}_\kv(t)=\hat{h}_k(t)
\cdot\vec{\psi}_\kv(t).
\end{eqnarray}

For a given condensate density $n_c(t)$ the solution for the evolution
operator $U(t)$ can be found exactly\cite{YRunpublished} and for a
slowly evolving $n_c(t)$ it is well approximated by an instantaneous
Bogoliubov transformation for $H_f^B(t)$,
\begin{eqnarray}
\label{UBt}
U(t)=
\begin{pmatrix}
u_k(t)e^{-i\int_0^t E_k(t')}&v_k(t)e^{i\int_0^t E_k(t')}\\
v_k(t)e^{-i\int_0^t E_k(t')}&u_k(t)e^{i\int_0^t E_k(t')}\\
\end{pmatrix},
\end{eqnarray}
where
\begin{eqnarray}
\label{ut}
u_k(t)&=&\sqrt{\oh\left(\frac{\epsilon_k+g_fn_c(t)}{E_k(t)} + 1\right)},\\
\label{vt}
v_k(t)&=&-\sqrt{\oh\left(\frac{\epsilon_k+g_fn_c(t)}{E_k(t)} - 1\right)},
\end{eqnarray}
and $E_k(t)=\sqrt{\epsilon_k^2+2g_fn_c(t)\epsilon_k}$.  Using this and
the relation of the atomic operators at the initial time to pre- and
post-quench Bogoliubov operators
\begin{equation}
\label{eq14}
U(0^-)\begin{pmatrix}\alpha_k\\\alpha^{\dagger}_{-k}\end{pmatrix}
=\begin{pmatrix}a_k(0^+)\\a^{\dagger}_{-k}(0^+)\end{pmatrix}
= U(0^+)\begin{pmatrix}\beta_k\\\beta^{\dagger}_{-k}\end{pmatrix},
\end{equation} 
we have
\begin{equation}
\label{a_t_relation}
\begin{pmatrix}
a_k(t)\\a^{\dagger}_{-k}(t)
\end{pmatrix}
= U(t)U^{-1}(0^+)U(0^-)
\begin{pmatrix}
\alpha_k\\\alpha^{\dagger}_{-k}
\end{pmatrix}.
\end{equation} 
We can now compute arbitrary dynamic atomic correlators, such as the
structure function, the rf spectroscopy signal, and the momentum
distribution function in terms of these time-dependent
matrices\cite{YRunpublished}. Focusing on the momentum distribution
function (measured in the JILA experiments\cite{Makotyn}), we find (see the Appendix)
\begin{eqnarray}
\label{n_k1}
  n_k(t)&=&\langle 0^-|a^\dagger_\kv(t)a_\kv(t)|0^-\rangle
,\nonumber\\
  &=&
\frac{\hat k^4 + \hat k^2(\sigma+1+\hat{n}_c)+2\sigma\hat{n}_c
+2\hat{n}_c(1-\sigma)
\sin^2\phi}
{2\hat k\sqrt{(\hat k^2+2\hat{n}_c)(\hat k^2+2\sigma)(\hat k^2+2)}},\nonumber\\
&&-\frac{1}{2}
\end{eqnarray}
where
\begin{eqnarray}
\label{phi}
\phi(\hat k,\hat t,\hat{n}_c(\hat t))&=&
\int_0^{\hat t} dt'\sqrt{\hat k^2(\hat k^2+2\hat{n}_c(t'))},
\end{eqnarray}
with normalized $\hat k\equiv k/\sqrt{8\pi n \tilde a_f}$,
$\hat{n}_c(\hat t)\equiv n_c(\hat t/\tilde g_f n)/n$, $\sigma\equiv
a_0/\tilde a_f$, and $\hat t\equiv t \tilde g_f n$.

We close this equation by using it to self-consistently compute the
time-dependent condensate density $n_c(t)=n-1/V\sum_\kv n_k(t)\equiv
n-n_d(t)$. We solve numerically the corresponding equation for $\hat
n_c(\hat t)$
\begin{eqnarray}
\label{ncSCeqn}
1-\hat{n}_c(\hat t)=\hat{n}_d(\hat t)&=&8\sqrt{\frac{2}{\pi}}(n
\tilde a^3_f)^{1/2}
\int_0^\infty d\hat k \hat k^2 n_{\hat k}(\hat t,\sigma,
\hat n_c(\hat t)),\nonumber\\
\end{eqnarray}
with the solution for $n_d(t)$ illustrated in Fig.\ref{ndtFig}. 
\begin{figure}[htb]
\centering
\includegraphics[width=0.45\textwidth]{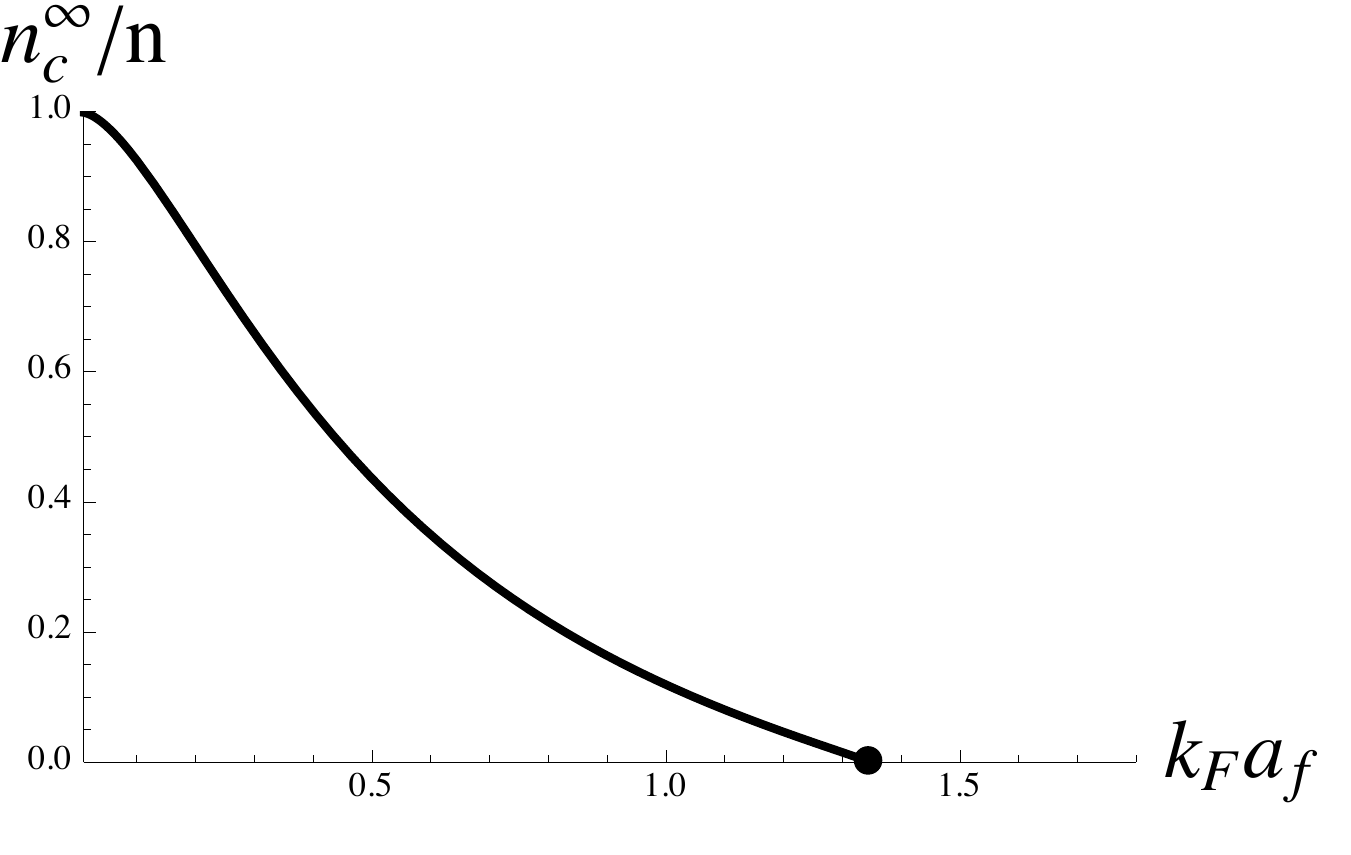}
\caption{(Color online) Long-time condensate fraction $n_c$ as a function of
  interaction strength, $k_n a_f$. For a sufficiently deep quench,
  $k_n a_f > 1.35$ (for $k_n a_0 = 0.01$), the asymptotic condensate
  density vanishes, suggesting a phase transition to a non-Bose
  condensed nonequilibrium steady state.}
\label{nc_Fig}
\end{figure}
In the long-time limit (averaging away the oscillatory component)
Eq.\ref{ncSCeqn} reduces to an equation for $n_c^\infty(k_na_f)$,
which then determines the steady-state momentum distribution function,
$n_k^\infty$ in Eq.\ref{nk_infty}.  The associated long-time depletion
$n_d^\infty$ deviates significantly from the ground-state value,
$n_d^{gs}=\frac{8}{3\sqrt{\pi}}\sqrt{n a_f^3}$ at the corresponding
$a_f$.
\begin{figure}[htb]
\centering
\includegraphics[width=0.45\textwidth]{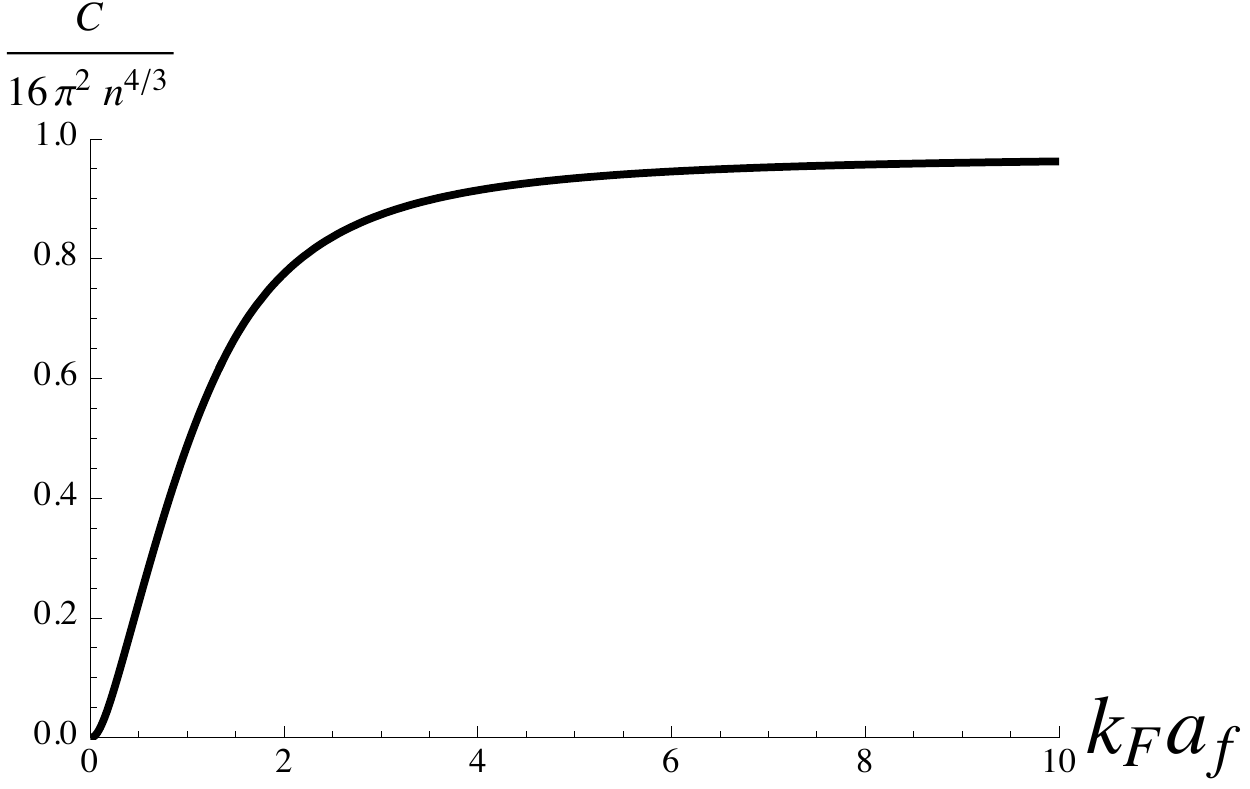}
\caption{(Color online) The effective contact $C(k_n a_f)$ associated with the
  asymptotic momentum distribution function as a function of the
  interaction strength $k_n a_f$ to which the system is quenched from
  $k_n a_0 = 0.01$.}
\label{C_Fig}
\end{figure}
As is clear from the inset of Fig.\ref{nk_ssFig}, the long-time
momentum distribution function $n_k^\infty$ exhibits a large momentum
$1/k^4$ tail, $n_{k\rightarrow\infty}^\infty = C/k^4$. We find that
the corresponding nonequilibrium contact, $C$ \cite{Tan,Wild,Braaten} is given
by
\begin{eqnarray}
C&=&16\pi^2k_n^4[(k_n\tilde{a}_f-k_na_0)^2
+(k_n\tilde{a}_f\frac{n_c^\infty}{n})^2]
\label{Cinfty}
\end{eqnarray}
and is illustrated in Fig.\ref{C_Fig}.

Finally, we observe that our long-time solution $n_c^\infty$ of
Eq.\ref{ncSCeqn} monotonically decreases with $k_n a_f$, vanishing at
the critical quench value of $k_n a_{fc}=1.35$ (see Fig.\ref{nc_Fig}). This
suggests\cite{commentTransition} that such deep quenches excite the
zero-temperature Bose gas to high enough energies, $E_{exc}$, so as to
fully deplete the condensate and to drive a nonequilibrium transition
to a non-BEC state. This is not an unreasonable nonequilibrium
counterpart of a thermal BEC-to-normal gas transition.

To summarize, we studied the dynamics of a resonant Bose gas,
following a deep quench to a large positive scattering
length. Utilizing a self-consistent extension of a Bogoliubov theory,
which allows us to approximately account for a large depletion and a
time-dependent condensate density, we computed the nonequilibrium
momentum distribution function, $n_k(t)$ and a variety of properties
derived from it. They show reasonable qualitative agreement with
recent experiments\cite{Makotyn}, but also leave many interesting open
questions for future studies\cite{YRunpublished}. These include an exact numerical solution of the self-consistent Eqs. \eqref{n_k1}-\eqref{ncSCeqn} to verify our approximate solution, and the
incorporation of the molecular state expected to appear on the
bound-state side of the resonance, through the analysis of the
two-channel model. Finally, research is under way to extend our work
to include quasi-particle interaction, which will account for the
thermalization expected on general grounds and observed
experimentally.

We thank P. Makotyn, D. Jin, and E. Cornell for sharing their data
with us before publication, and acknowledge them, A. Andreev, D. Huse,
V. Gurarie, and A. Kamenev for stimulating discussions. This research
was supported by the NSF through DMR-1001240.

Note added: 
A complementary analysis of the experiment in \cite{Makotyn} has been recently posted \cite{Sykes}. It utilizes a time-dependent variational approach \cite{Zhou}, which is expected to be equivalent to our self-consistent Bogoliubov theory

\section{Appendix}
\renewcommand{\theequation}{A\arabic{equation}}    
\setcounter{equation}{0}  

In this appendix we fill in some of the technical details leading to the
results reported in the main text. Starting with Eqs. \eqref{UBt}-\eqref{vt}, it is straightforward
to show
\begin{eqnarray}
U^{-1}(0^{+})U(0^{-})=\begin{pmatrix}\cosh\Delta\theta_k&\sinh\Delta\theta_k\\
\sinh\Delta\theta_k&\cosh\Delta\theta_k\end{pmatrix},
\end{eqnarray}
where
\begin{eqnarray}
\label{parameters}
\Delta\theta_k=\frac{1}{2}\cosh^{-1}(\frac{1}{2}(\frac{E_f}{E_i}+\frac{E_i}{E_f})),\;\;\;\epsilon_k=\frac{k^2}{2m}
\nonumber\\
E_i\equiv\sqrt{(\epsilon_k)^2+2\epsilon_k ng_i},\;\;\;E_f\equiv\sqrt{(\epsilon_k)^2+2\epsilon_k ng_f}.
\end{eqnarray}
Combining this with \eqref{a_t_relation} and \eqref{UBt}, we obtain
\begin{eqnarray}
\label{a_t_relation2}
\begin{pmatrix}a_k(t)\\ a^{\dagger}_{-k}(t)\end{pmatrix}=
\begin{pmatrix}u_k(t)e^{-i\int_0^t E_k(t')dt'}&v_k(t)e^{i\int_0^t E_k(t')dt'}\\
v_k(t)e^{-i\int_0^t E_k(t')dt'}&u_k(t)e^{i\int_0^t E_k(t')dt'}\end{pmatrix}
\nonumber\\
\times\begin{pmatrix}\cosh\Delta\theta_k&\sinh\Delta\theta_k\\
\sinh\Delta\theta_k&\cosh\Delta\theta_k\end{pmatrix}
\begin{pmatrix}\alpha_k\\ \alpha^{\dagger}_{-k}\end{pmatrix}\;\;\;\;
\end{eqnarray}
giving the evolution of atomic operators $a_k(t)$ and $a_k^\dagger(t)$
following the quench at time $t=0$. Taking the initial gas to be in thermal equilibrium, prior to the
quench, the quasi-particles $\alpha_k$ and $\alpha_k^\dagger$ obey the Bose-Einstein distribution
\begin{eqnarray}
\langle 0^-|\hat{\alpha}^{\dagger}_{k}\hat{\alpha}_k|0^-\rangle=\langle 0^-|\hat{\alpha}_k\hat{\alpha}^{\dagger}_{-k}|0^-\rangle-1=\frac{1}{e^{E_i/k_B T}-1}.
\end{eqnarray}
At $T=0$ of interest to us here, these as usual reduce to
\begin{eqnarray}
\langle 0^-|\hat{\alpha}^{\dagger}_{k}\hat{\alpha}_k|0^-\rangle=\langle 0^-|\hat{\alpha}_k\hat{\alpha}^{\dagger}_{-k}|0^-\rangle-1=0.
\end{eqnarray}
Combining these with \eqref{a_t_relation2}, we can straightforwardly evaluate the
atom momentum distribution function $n_k(t)$, at time $t$ after the quench,
obtaining
\begin{eqnarray}
\label{n_k2}
n_k(t)=\langle 0^-|a^\dagger_\kv(t)a_\kv(t)|0^-\rangle\quad\quad\quad\quad
\nonumber\\
=
|(u(t)e^{-i\int_0^t E_k(t')dt'}\sinh\Delta\theta_k\nonumber\\
+v(t)e^{i\int_0^t E_k(t')dt'}\cosh\Delta\theta_k)|^2.
\end{eqnarray}
Now using \eqref{ut} \eqref{vt} and \eqref{parameters} inside \eqref{n_k2}, we obtain our key result for
$n_k(t)$ reported in Eqs. \eqref{n_k1} and \eqref{phi} of the main text, with normalized parameters
$\hat k$, $\hat{n}_c(\hat t)$ and $\hat{t}$ defined there.


\begin{thebibliography}{99}
\bibitem{commentTransition} Because our self-consistent Bogoliubov
  analysis is uncontrolled we cannot be certain that this
  nonequilibrium transition is not an artifact of this
  approximation.
\end{thebibliography}
\end{document}